\documentclass[final,twocolumn,10pt]{asme2ej}

\usepackage{epsfig} 
\usepackage{amsmath} 
\usepackage{soul,color} 
\soulregister\cite7
\soulregister\ref7

\title{Buoyancy effect on the flow pattern and the  thermal performance of an
array of circular cylinders}

\author{Francesco Fornarelli\thanks{Corresponding author.} \\
    \affiliation{ Senior post-doc \\ 
Department of Mechanics, \\Mathematics and Management \\
Polytechnic of Bari, \\via Orabona 4, 70125 Bari, Italy \\
INFN sez. Lecce, 73100 Lecce, Italy\\
        Email: francesco.fornarelli@poliba.it
    }
}

\author{Antonio Lippolis \\
    \affiliation{ Full Professor \\ 
Department of Mechanics, \\Mathematics and Management \\
Polytechnic of Bari, \\via Orabona 4, 70125 Bari, Italy \\
        Email: antonio.lippolis@poliba.it
    }
}
\author{Paolo Oresta \\
    \affiliation{ Assistant Professor \\ 
Department of Mechanics, \\Mathematics and Management \\
Polytechnic of Bari, \\via Orabona 4, 70125 Bari, Italy \\
INFN sez. Lecce, 73100 Lecce, Italy\\
        Email: paolo.oresta@poliba.it
    }
}

\begin{document}

\maketitle    

\begin{abstract} {\it In this paper we found, by means of numerical
simulations, a transition in the oscillatory character of the flow field for a
particular combination of buoyancy and spacing in an array of six circular
cylinders at a Reynolds number of $100$ and Prandtl number of $0.7$. The
cylinders are iso-thermal and they are aligned with the {\it Earth}
acceleration ($\bf g$). According to the array orientation, an aiding or an
opposing buoyancy is considered. The effect of natural convection with respect
to the forced convection is modulated with the Richardson number, $Ri$, ranging
between $-1$ and $1$. Two values of center to center  spacing ($s=3.6d - 4d$)
are considered. The effects of buoyancy and spacing on the flow pattern in the
near and far field are described. Several transitions in the flow patterns are
found and a parametric analysis of the dependence of the force coefficients and
Nusselt number with respect to the Richardson number is reported. For $Ri=-1$,
the change of spacing ratio from $3.6$ to $4$ induces a transition in the standard
deviation of the force coefficients and heat flux. In fact the transition
occurs due to rearrangement of the near field flow in a more ordered wake
pattern.  Therefore, attention is focused on the influence of geometrical and
buoyancy parameters on the heat and momentum exchange and their fluctuations.
The available heat exchange models for cylinders array provide a not accurate
prediction of the Nusselt number in the cases here studied.
}
\end{abstract}

\begin{nomenclature}
\entry{$C_d,\, C_l$}{drag and lift coefficients}
\entry{$F_x,\, F_y$}{force components, $N$}
\entry{$H$}{heat transfer coefficient, $W/(m^2 K)$}
\entry{$T^\star$}{temperature, $K$}
\entry{$T^\star_L$}{inflow temperature, $K$}
\entry{$T^\star_H$}{cylinder temperature, $K$}
\entry{$U^\star$}{inflow velocity, $m/s$}
\entry{$d$}{cylinder diameter, $m$}
\entry{$f$}{frequency, $s^{-1}$}
\entry{$\bf{g}$}{Earth acceleration, $m/s^2$}
\entry{$k$}{fluid thermal conductivity, $W/(m K)$}
\entry{$p$}{dimensionless pressure}
\entry{$q$}{transversal cylinder spacing, $m$}
\entry{$s$}{in-line cylinder spacing, $m$}
\entry{$\bf{u}$}{dimensionless fluid velocity vector}
\noindent
\textbf{Greek Letters}

\noindent
\entry{$\Delta$}{cylinder-inflow temperature difference, $K$; $(T^\star_H - T^\star_L)$}
\entry{$\alpha$}{thermal expansion coefficient, $K^{-1}$}
\entry{$\kappa$}{thermal diffusivity, $m^2/s$}
\entry{$\nu$}{kinematic viscosity, $m^2/s$}
\entry{$\rho$}{fluid density, $kg/m^3$}
\noindent
\textbf{Dimensionless Numbers}

\noindent
\entry{$Gr$}{Grashof number; $g \alpha \Delta d^3 / \nu^2$}
\entry{$Nu$}{Nusselt number; $H d /k$}
\entry{$Pr$}{Prandtl number; $\nu/\kappa$}
\entry{$Re$}{Reynolds number; $U^\star d/\nu$}
\entry{$Ri$}{Richardson number; $Gr/Re^2$}
\entry{$St$}{Strouhal number; $f d/U^\star$}
\end{nomenclature}

\section{Introduction}

The flow around multiple bluff bodies is a prototype of many engineering
problems ranging from heavy-duty to micro-devices applications. Offshore
pipelines, electrical power lines, electronic and bio-tech devices are just few
examples of applications in which flow interacts with multiple bluff bodies.
Among them the heat exchangers involve a wide range of engineering
applications.  In general, they consist of solid surfaces at a certain
temperature immersed in a cross flow at a different temperature. In particular
tube bundles heat exchangers are common in several micro applications such as
in heat exchange control in Li-ion batteries~\cite{Ye} or in biomedical
devices.  For instance, the thermal performance of lab-on-chip devices assumes
a key role in a wide range of biological applications, such as the study of
tumor cells under constant temperature~\cite{Wan2013}.  The small dimensions
and the low flow velocity induce unsteady laminar
regimes~\cite{PhysRevE.81.036305}.  The oscillations induced by the flow
patterns affects the force and thermal response of such devices that have to be
taken into account in the design
process~\cite{Duryodhan2016,Selimefendigil2014}.  Indeed the prediction of the
performance of these devices is still the subject of study.  Numerical
simulation of the flow field and heat exchange aids to give a detailed overview
of the flow quantities involved in such a flow.  In the present study the flow
field around six circular cylinders has been investigated by means of numerical
simulations.  Three dimensionless parameters are involved in this type of
problem: the Reynolds ($Re$), Prandtl ($Pr$) and Richardson ($Ri$) number
defined as: 
\begin{eqnarray} Re=\frac{U^\star d}{\nu} \quad & Pr=\frac{\nu}{\kappa} \quad &
Ri=\frac{Gr}{Re^2} \end{eqnarray} 
where $U^\star$, $d$, $\nu$ and $\kappa$ are , respectively, the inflow
velocity, the cylinder diameter, the kinematic viscosity of the fluid and its
thermal diffusivity. $Gr = g \alpha \Delta d^3 / \nu^2$ is the Grashof number
where $g$, $\alpha$ and $\Delta$ are, respectively, the {\it Earth}
acceleration modulus, thermal expansion coefficient and the temperature
difference between the cylinder surface and the cross flow free stream
temperature.  The Richardson number represents the importance of the natural
convection with respect to the forced convection. Usually the range in which
both effects are present is characterized by values of $-1\le Ri\le 1$ and it
is called mixed convection.  The higher is the absolute value of the Richardson
number the smaller is the effect of the convection forced by the inlet velocity
with respect to the natural convection.  Forced convection, $Ri = 0$, is the
first step to study the heat exchange between the bluff bodies and the flow.
In this case the flow is not influenced by temperature in the hypothesis of
small temperature differences between the bluff bodies and the flow temperature
with respect to the dominant velocity convection.  In Fornarelli et
al.~\cite{FornarelliOresta} the authors investigated the flow field and the
heat exchange around six circular cylinders by means of numerical simulation.
The tests have been done in case of forced convection ($Ri=0$), and a
transition in the flow patterns and in the heat exchange has been identified.
The  flow pattern transition occurred for a spacing ratio between $3.6$ and
$4$. The flow is unsteady and the heat exchange of each cylinder is strongly
influenced by the vorticity dynamics.  The influence of the buoyancy force on
the flow field is expected to be important in order to change the flow and heat
transfer dynamics.  The buoyancy force influences both the near and the far
field with respect to a solid obstacle immersed in a flow affecting the
boundary layer separation and the onset of the vortex shedding in the
wake~\cite{Chatterjee2014152,Clifford2014}.  In a two cylinders configuration,
mixed convection with aided buoyancy, aligned to the free stream velocity, has
a stabilizing effect on the flow pattern, vice versa the opposed buoyancy
anticipates the boundary layer separation at the cylinder surface and makes the
flow more unstable~\cite{Patnaik2000560}.  Nevertheless a simple two bodies
model is not able to predict the multiple cylinders configuration behaviour
because of a more complex wake interference phenomenon that affects the
downstream cylinders.  In literature, the in-line configuration of multiple
heated cylinders considering the effect of buoyancy force has not been
extensively investigated.  Khan et al.~\cite{Khan20064831} studied the thermal
response of isothermal tube bundle of circular cylinders in in-line and
staggered configuration over a wide range of Reynolds number but only in case
of forced convection ($Ri=0$) using an integral solution of the boundary layer
equations.  Multiple row configurations have been studied focusing on the
characterization of the mean value of the force and the heat transfer
coefficients~\cite{Wang2000819, Gowda19981613}.  The analytical results are
able to model a wide parameter range in the hypothesis of an infinite number of
rows, but the unsteady characteristics cannot be extrapolated. Moreover at a
Reynolds number of $100$ the effects of wake interference on the heat exchange
is not easily predictable by means of simplified models~\cite{Zukauskas197293,
Gnielinski1975145}.  The aim of the present work is to shed light on  the
oscillatory characteristic of the force and heat transfer coefficients in case
of a single in-line array of six circular cylinders.  In order to retain
the two-dimensional character of the flow field, our simulations have been
carried out at $Re = 100$, being in literature, for the case of a single
cylinder, $Re = 200$ the threshold for the transition from two to
three-dimensional flow~\cite{Barkley1996215} and also for two identical in-line
cylinders, in a wide range of in-line spacings, the two dimensional character
of the flow is retained at $Re=100$, as reported in the works of Carmo et
al.~\cite{FLM:7244044,Carmo20101}. They state that the onset of
three-dimensional instabilities, for a spacing ratio between $3.6$ and $4$,
occurs for $Re\simeq150$.   Three-dimensional instabilities induced by the
buoyancy force are limited being the buoyancy force modulated in the range $-1
< Ri < 1$~\cite{Maas20033069,Ren20043103,Boirlaud201282}. A detailed
description of the flow patterns and temperature distribution have been
reported.  Moreover the dependence of the dimensionless force and heat transfer
coefficients ($C_d, C_l, Nu$) have been reported with a quantitative analysis
of their mean and oscillating components. 

\section{Numerical setup} 

\begin{figure} \centering
\includegraphics[width=0.99\linewidth]{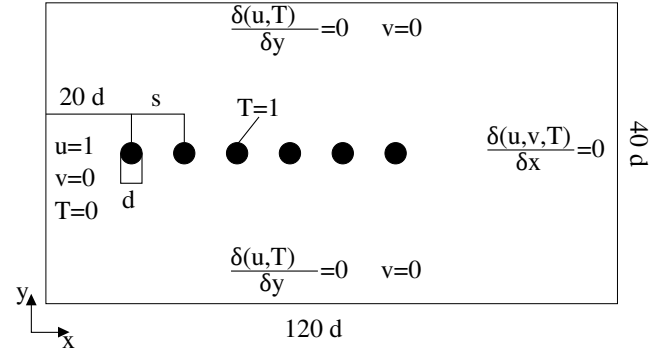} \caption{Outline
of the numerical setup}\label{fig:bc-sketch} \end{figure}

The incompressible two-dimensional Navier-Stokes equations and the heat
transfer equation are considered. Here follows the governing equations in
dimensionless form:
\begin{equation}\label{eq:NS1} \frac{\partial{\mathbf u}}{\partial t}+{\mathbf
u}\nabla {\mathbf u} = -\nabla p +\frac{1}{Re} \nabla^2 {\mathbf u}+Ri\, T
\end{equation} 
\begin{equation}\label{eq:NS2} \nabla \cdot {\mathbf u} = 0.  \end{equation}  
\begin{equation}\label{eq:Energy} \frac{\partial T}{\partial t}+\mathbf u
\nabla T= \frac{1}{Re Pr} \nabla^2 T \end{equation} 
where lengths are scaled by the cylinder diameter ($d$) and the velocities by
the free-stream velocity ($U^\star$). The temperature is scaled with respect to
the free-stream temperature ($T^\star_L$) and the constant temperature of the
cylinders ($T^\star_H$) as follows: 
\begin{equation} T=(T^\star - T^\star_L)/(T^\star_H - T^\star_L).
\end{equation}
In the momentum equation (~\ref{eq:NS1}), in order to take into account the
buoyancy force,  the Boussinesq approximation has been considered . $Ri$
represents the ratio between the buoyancy and the inertial force. The gravity
vector $\mathbf g$ is aligned with the streamwise direction $x$.  The Reynolds
number and the Prandtl number have been kept fixed, $Re=100$ and $Pr=0.7$,
respectively.  Direct numerical simulations have been performed using a
fractional step projection method to enforce the continuity equation with a
pressure correction approach \cite{popinet2003}. The advection terms are
treated by means of a Godunov procedure using a second order upwind method. For
the viscous terms an implicit Crank-Nicholson scheme has been implemented. The
Poisson equation for the pressure correction step is iterated until the
local error on the continuity equation is greater than $10^{-6}$.  In
fig.~\ref{fig:bc-sketch} a schematic representation of the computational domain
and boundary conditions is shown.  The numerical domain is $120\,d$ long in the
streamwise direction and $40\,d$ wide in the transversal direction.  An adaptive
mesh refinement approach has been implemented in the numerical
code~\cite{popinet2003}. The refinement strategies include the limiting of the
velocity and temperature difference between two adjacent grid cells every
time-step. A detailed and comprehensive grid independence study and a
sensitivity analysis on the domain dimension has been carried out as reported
in a previous work of the authors~\cite{FornarelliOresta}.  At the inflow
uniform flow is imposed with $u=1$, $v=0$ and $T=0$, while at the outflow the
spatial variation of the velocity components and temperature in the streamwise
direction are imposed as follows: $\partial u/\partial x=0$, $\partial
v/\partial x=0$ and $\partial T/\partial x=0$.  Symmetry boundary conditions
are imposed on the walls ($v=0$, $\partial u/\partial y=0$ and $\partial
T/\partial y=0$).  The cylinders surfaces are kept at a constant dimensionless
temperature equal to $1$ whereas the inlet flow temperature is $0$. The spacing
between the cylinders is constant and two different values are considered,
$s=3.6d$ and $s=4.0d$.  All the quantities reported in the results section are
averaged over $1000$ dimensionless time units ($t=d/U^\star$) in order to
achieve the convergence of the statistics. To ensure initial condition
independence of the results statistics are collected after $800\,t$.  
The numerical results of the force coefficients and heat transfer coefficient
are reported. Here, $C_d$ is the drag coefficient, $C_l$ is the lift
coefficient and $Nu$ is the Nusselt number, which are defined as:
\begin{equation} C_d = \frac{F_x}{0.5\rho {U^\star}^2 d}, \qquad C_l =
\frac{F_y}{0.5\rho {U^\star}^2 d}, \qquad Nu = \frac{H d}{k}, \end{equation} 
where $F_x$ and $F_y$ are the drag and lift force per unit length, respectively.
$H$ is the local heat transfer coefficient and $k$ is the thermal conductivity.

\section{Results}

\subsection{Effect of spacing and buoyancy force on temperature spatial
distribution}\label{subsec:T-distribution}

\begin{figure*}
\includegraphics[width=0.95\linewidth]{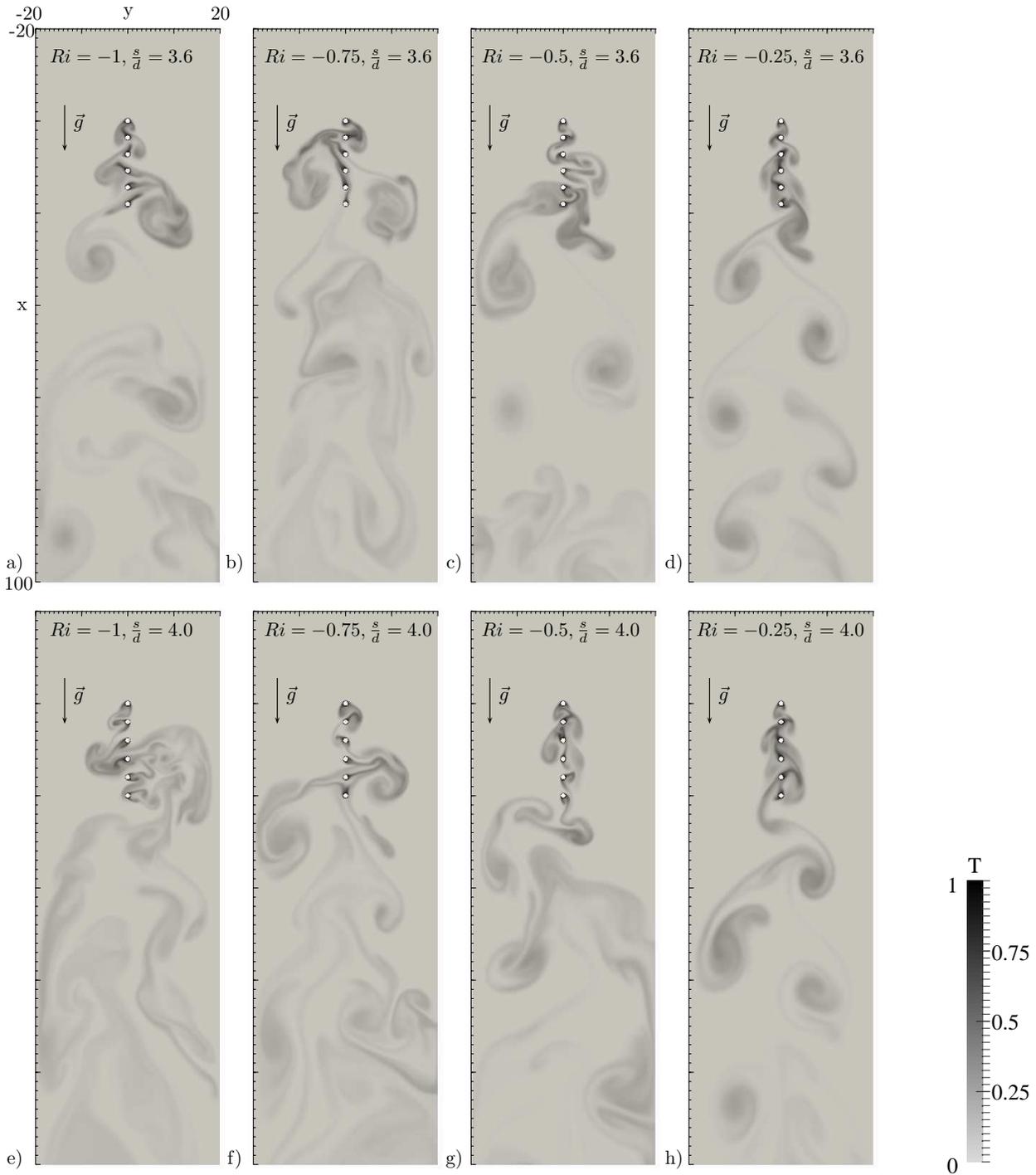}
\caption{Comparison of the dimensionless temperature distribution at $t=1800$
for opposing buoyancy $Ri<0$. The domain is placed in vertical position with
the free stream velocity oriented downward.} \label{fig:T-contour-neg}
\end{figure*}

\begin{figure*}
\includegraphics[width=0.95\linewidth]{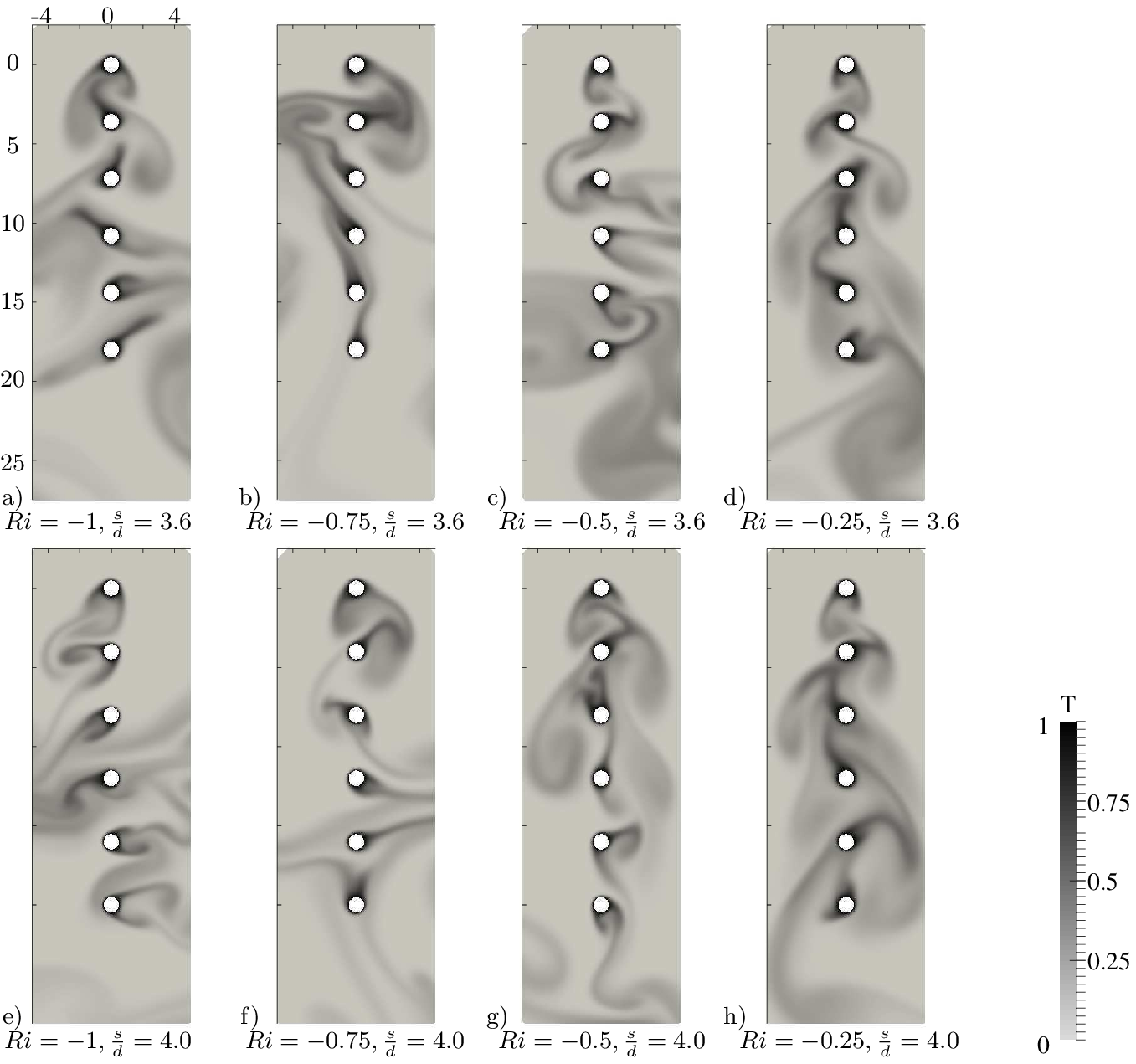}
\caption{Near field comparison of the dimensionless temperature distribution at
$t=1800$ for opposing buoyancy $Ri<0$. The domain is placed in vertical
position with the free stream velocity oriented downward.}
\label{fig:T-contour-neg-zoom} \end{figure*}

\begin{figure*}
\includegraphics[width=0.95\linewidth]{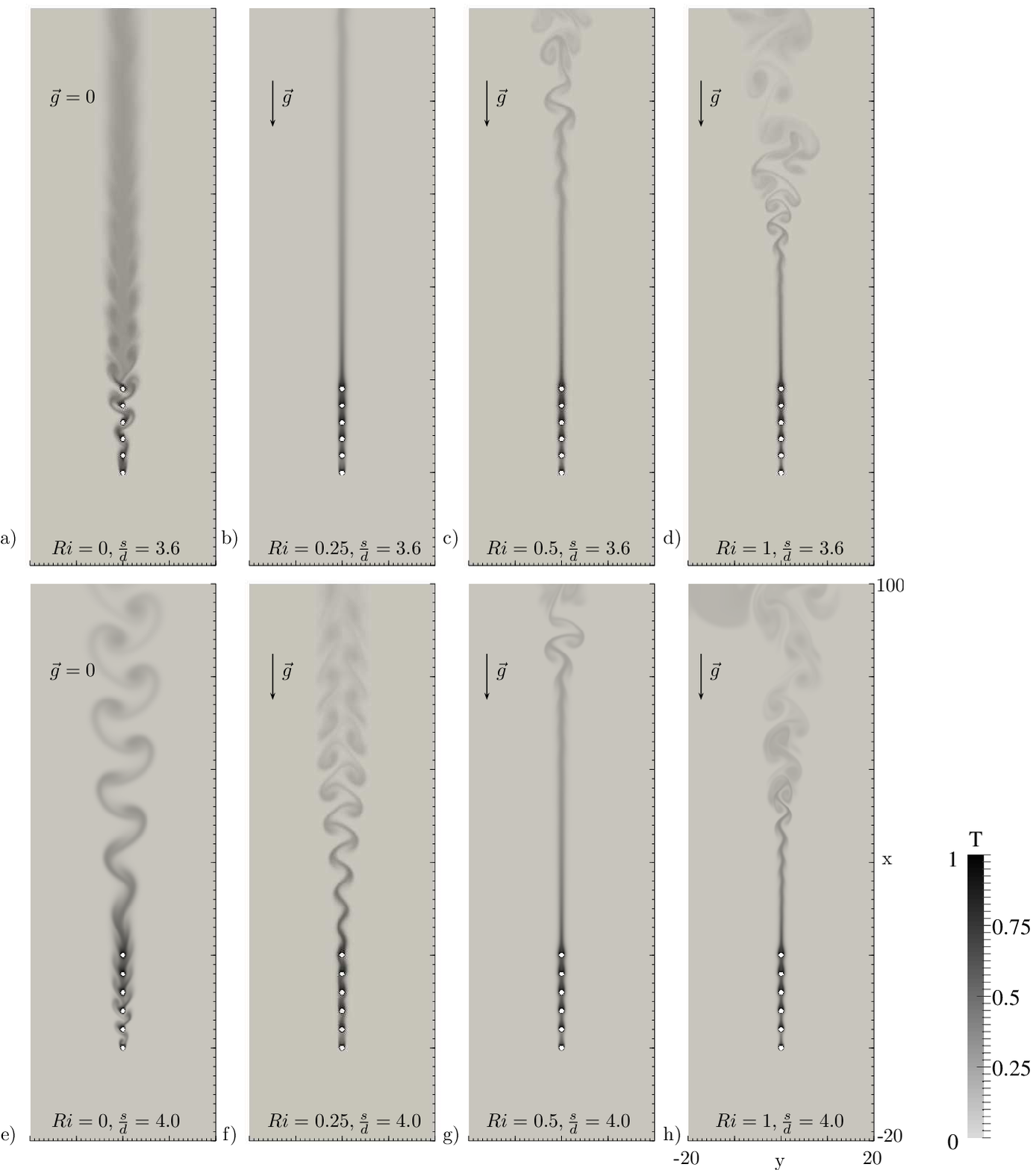}
\caption{Comparison of the dimensionless temperature distribution at $t=1800$
for aiding buoyancy $Ri\ge0$. The domain is placed in vertical position with
the free stream velocity oriented upward.} \label{fig:T-contour-pos}
\end{figure*}

\begin{figure*}
\includegraphics[width=0.95\linewidth]{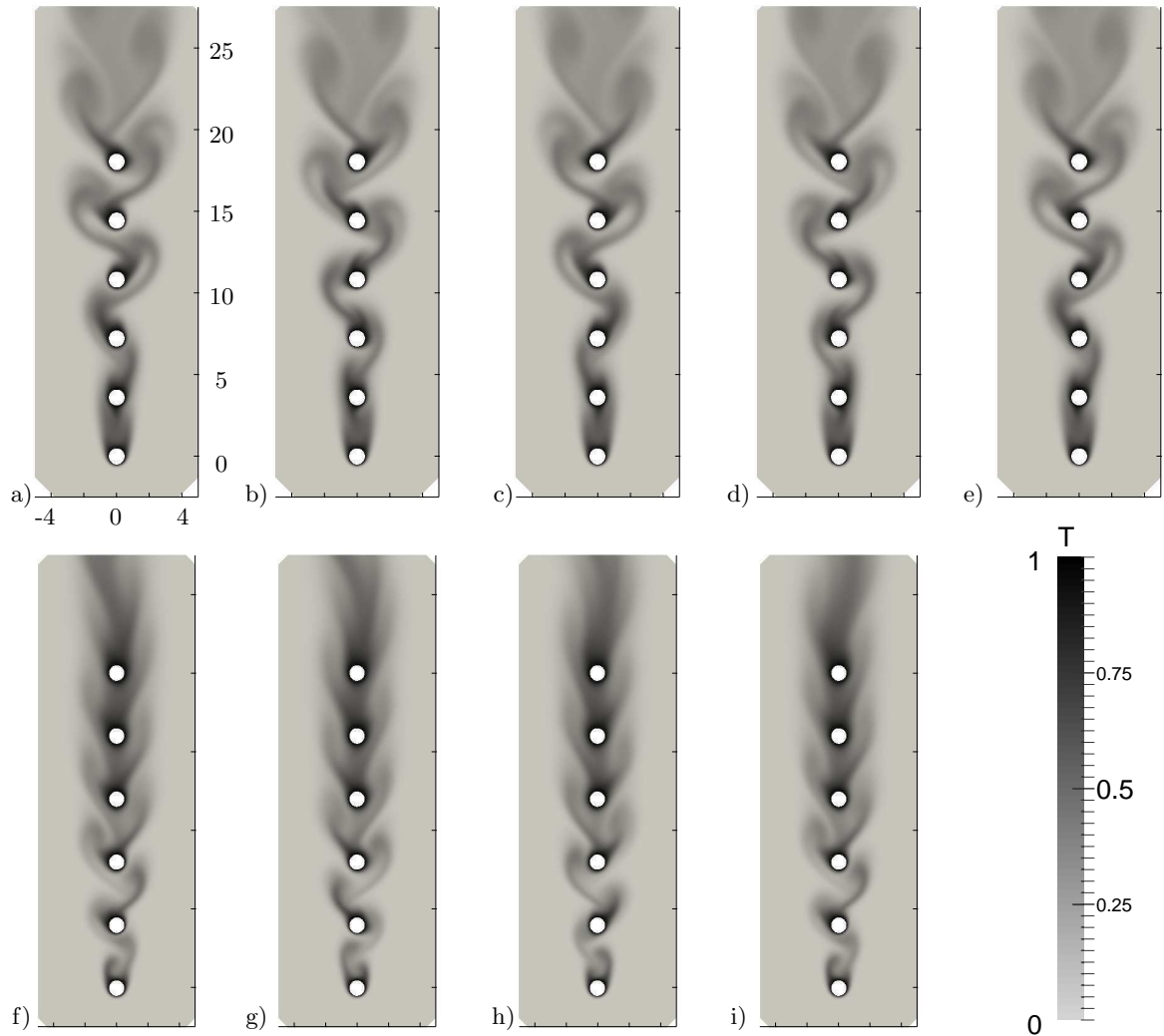}
\caption{Snapshots of the dimensionless temperature distribution over one
oscillation cycle in case of forced convection $Ri=0$, for $s/d=3.6$ at
$t=1800.0$ (a), $t=1802.5$ (b), $t=1805.0$ (c), $t=1807.5$ (d), $t=1810.0$ (e);
and $s/d=4.0$ at $t=1800.0$ (a), $t=1802.5$ (b), $t=1805.0$ (c), $t=1807.5$
(d). The domain is placed in vertical position with the free stream velocity
oriented upward.} \label{fig:timeseries-Ri0} \end{figure*}

  The flow and thermal behaviour of six in-line cylinders configuration as
function of the Richardson number for two values of spacing ratio are
investigated. First, a qualitative comparison of the temperature distribution
contours at the same instant is reported. In fig.~\ref{fig:T-contour-neg}
opposing buoyancy cases ($Ri<0$) are shown; in these cases, due to the complex
spatial distribution of the temperature contours, a close up of the near field
is also reported in fig.~\ref{fig:T-contour-neg-zoom}. In
fig.~\ref{fig:T-contour-pos} forced convection and aiding buoyancy cases
($Ri\ge0$) are shown. The figures are oriented according to the direction of
the gravity vector, $\mathbf{g}$. Thus, the free stream velocity is oriented
downward or upward for negative or positive values of $Ri$, respectively. The
counter-oriented buoyancy induces a thermal wake widening, instead of the
aiding buoyancy cases, where the wake appears narrower.  This phenomenon
affects both the spacings remarking that the buoyancy influences the boundary
layer separation around bluff obstacles as already described in previous
analogous studies~\cite{Patnaik2000560,Chatterjee2014152}. In case of forced
convection, $Ri=0$, the $s/d=3.6$ configuration does not show a secondary
instability of the wake, whereas at $s/d=4$ the wake evolves in a meandering
configuration as described in Fornarelli et al.~\cite{FornarelliOresta}
(fig.~\ref{fig:T-contour-pos}a-e). This behaviour is related to the flow
structures shed by the cylinders. In fig.\ref{fig:timeseries-Ri0} the
temperature distributions of about one oscillating cycle at $Ri=0$
for both the spacings are reported. At $s/d=3.6$
(fig.\ref{fig:timeseries-Ri0}a-e) a phase shift in the shedding behaviour of
two successive cylinders produces alternate shedding structures on the right
and left hand side with respect to the array centerline.  Whereas, at $s/d=4.0$
(fig.\ref{fig:timeseries-Ri0}f-i) the flow structures are shed in phase.  In
case of mixed convection, the opposing buoyancy widens the wake according to
the enhanced boundary layer separation at the cylinders surface.  Indeed for
$Ri\neq0$ the buoyancy force adds its contribution as source term in the
momentum equation, eq.~(\ref{eq:NS1}), affecting the streamwise component of
the velocity, especially near the cylinders, where the temperature is the
highest.  Being the cylinders the only source of vorticity, the flow patterns
and the temperature distribution are very sensitive to $Ri$.  In particular for
$Ri<0$ the opposing buoyancy force induces the boundary layer instabilities in
the near field as reported in detail in fig.\ref{fig:T-contour-neg-zoom}.
It is worth to note that for both the spacings, in the cases of $Ri\le-0.5$
(fig.~\ref{fig:T-contour-neg}-a,b,c,e,f,g), the opposing buoyancy affects the
vortex shedding of the array. The more is the opposing buoyancy the more
chaotic the temperature pattern appears, except for the case with $s/d=3.6$ at
$Ri=-1$ (see fig.~\ref{fig:T-contour-neg}-a).  Although at $Ri=-1$ a more
chaotic configuration is expected, at $s/d=3.6$, enhancing the opposing
buoyancy from $Ri=-0.75$ to $Ri=-1$, a well-ordered wake pattern can be
recognized, due to a reorganization of the vorticity interaction in the near
field of the array.  Hence, consequences on the performance of the array in
terms of force and heat exchange are expected and will be detailed below.  On
the other hand the aiding buoyancy, $Ri>0$, stabilizes the flow field.  With
respect to the case of forced convection the wake is narrower. The flow field
around the cylinders appears very stable and no shedding structures appear
behind the array. For $Ri=0.25$ and $s/d=3.6$ the temperature distribution in
the flow field reveals a stable wake pattern without any secondary instability,
even in the far field (see fig.~\ref{fig:T-contour-pos}-b).  The increasing of
the buoyancy force triggers a Kelvin-Helmholtz instability of the wake in the
far field. For $s/d=4$ the increasing of $Ri$ from $0$ to $0.25$ is not
sufficient to suppress the wake oscillation (see
fig.~\ref{fig:T-contour-pos}-f).  Increasing the buoyancy force at $Ri=0.5$,
the wake oscillation is suppressed and there is only a secondary oscillation in
the far field  (see fig.~\ref{fig:T-contour-pos}-g).

\subsection{Drag and lift forces analysis}

\begin{figure} \includegraphics[width=0.99\linewidth]{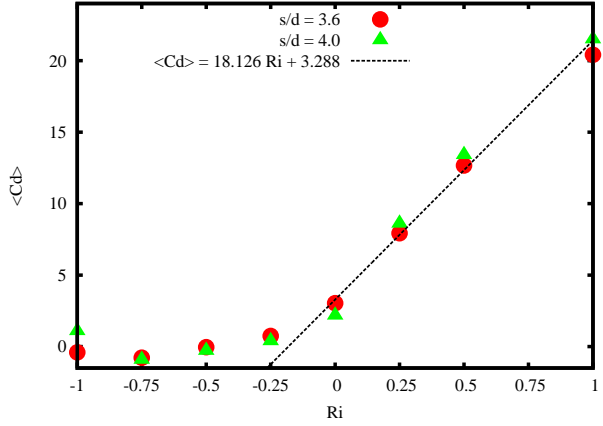}
\caption{Time averaged drag coefficient ($\langle C_d \rangle$) of the array
with respect to the Richardson number ($Ri$) for $s/d=3.6$ and $s/d=4.0$. The
linear fitting of the aiding buoyancy cases ($0\le Ri\le1$) is
reported.}\label{fig:Cd-Ri} \end{figure}

\begin{figure*}
\includegraphics[width=0.99\linewidth]{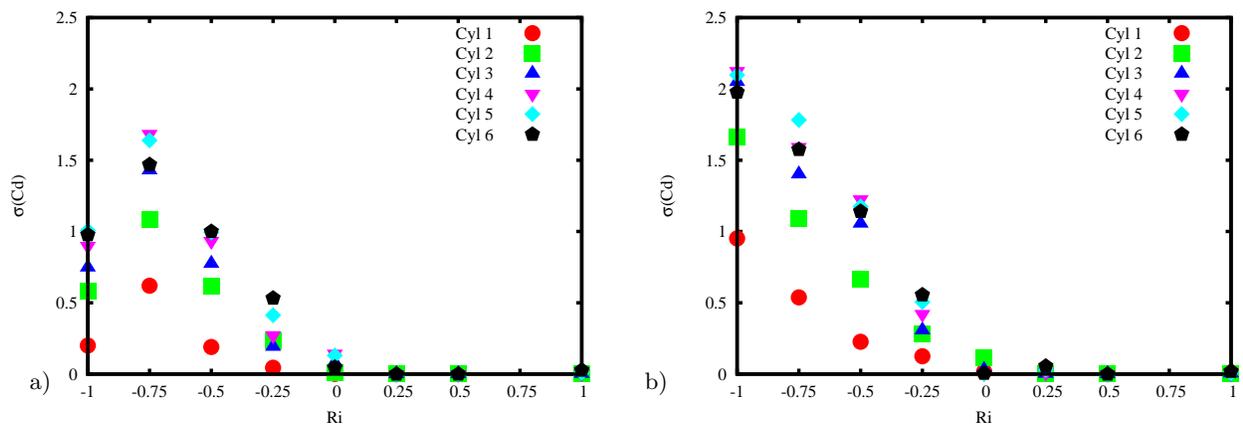}
\caption{Standard deviation of the drag coefficient ($\sigma(C_d)$) of each
cylinder with respect to the Richardson number ($Ri$) for $s/d=3.6$ (a) and
$s/d=4.0$ (b).}\label{fig:sigmaCd-Ri-sxx} \end{figure*}

\begin{figure} \includegraphics[width=0.99\linewidth]{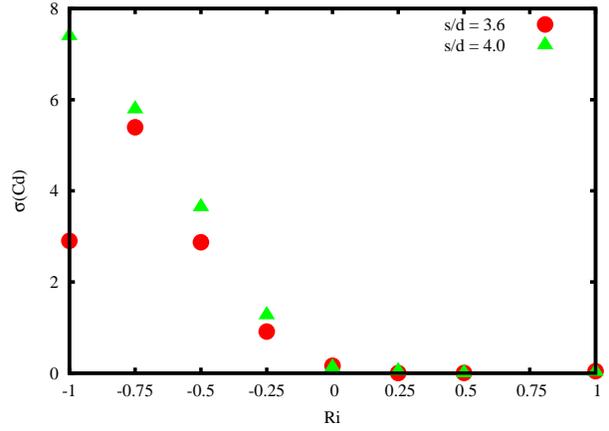}
\caption{Standard deviation of the drag coefficient ($\sigma(C_d)$) averaged
over six cylinders with respect to the Richardson number ($Ri$) for $s/d=3.6$
and $s/d=4.0$.}\label{fig:sigmaCd-Ri} \end{figure}

\begin{figure} \includegraphics[width=0.99\linewidth]{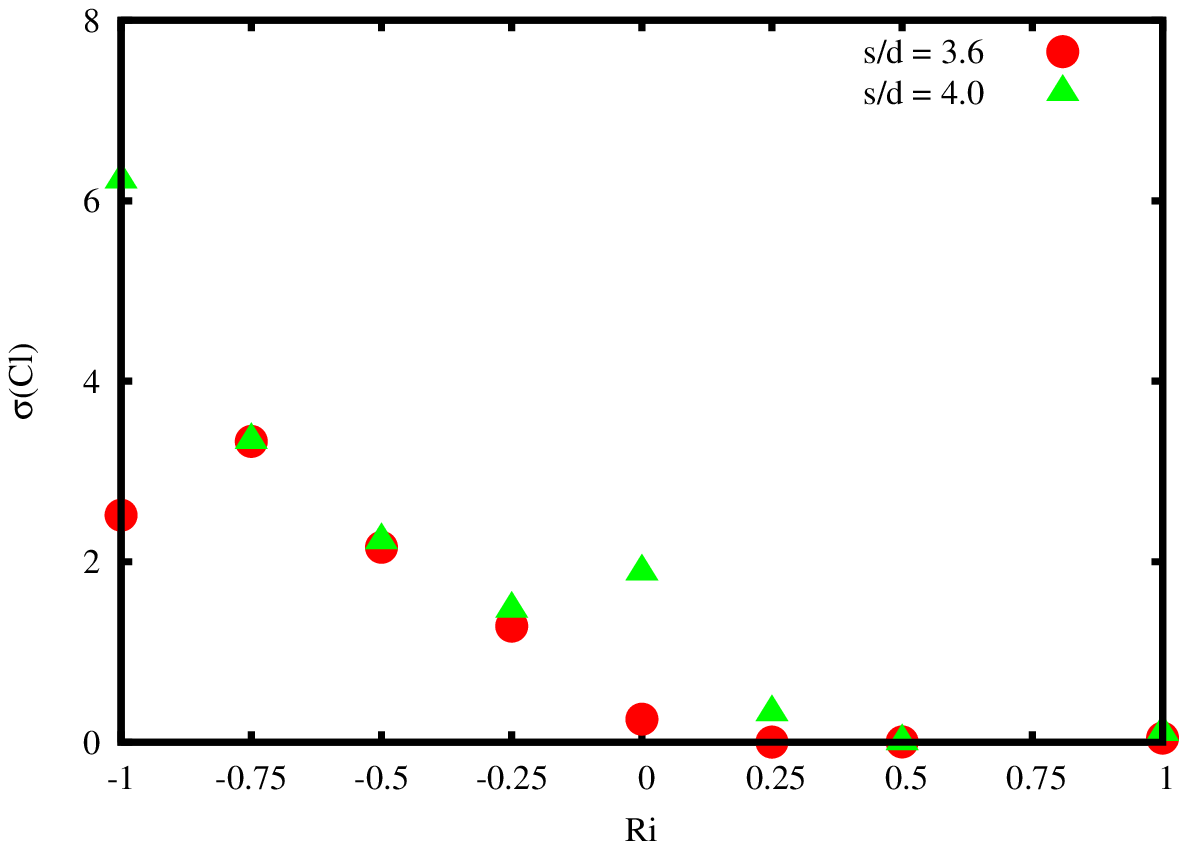}
\caption{Standard deviation of the lift coefficient ($\sigma(C_l)$) averaged
over six cylinders with respect to the Richardson number ($Ri$) for $s/d=3.6$
and $s/d=4.0$.}\label{fig:sigmaCl-Ri} \end{figure}

\begin{figure}
\includegraphics[width=0.99\linewidth]{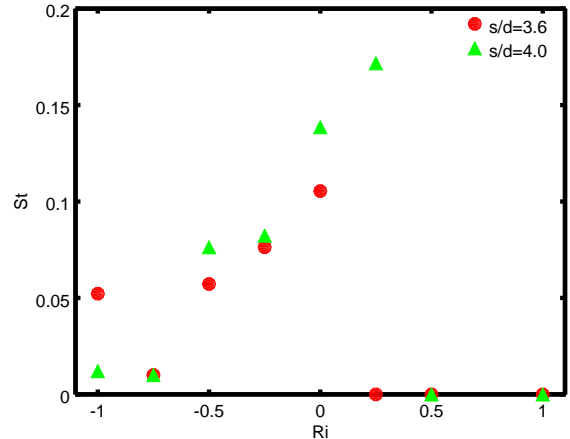}
\caption{Strouhal number ($St$) of the first cylinder with respect to the
Richardson number ($Ri$) for $s/d=3.6$ and $s/d=4.0$.}\label{fig:St-Ri}
\end{figure}

The drag coefficient of the entire array, $C_d (t)=\sum_{i=1}^6 {C_d}_i (t)$
has been measured for each value of $Ri$ and $s/d$. Being $C_d$ a function of
time, its time average, $\langle C_d \rangle$, and standard deviation,
$\sigma(C_d)$, have been calculated. In fig.~\ref{fig:Cd-Ri} the time averaged
drag coefficient is reported. The opposing buoyancy reduces the drag of the
array with respect to the force convection case for both the spacings. At
$Ri=-0.75$, a minimum time average drag coefficient is found, reaching negative
values, $\langle C_d \rangle=-0.778$ for $s/d=3.6$ and $\langle C_d
\rangle=-0.914$ for $s/d=4$. Indeed, at $Ri=-1$, the drag coefficient are
$\langle C_d \rangle=-0.401$ at $s/d=3.6$ and $\langle C_d \rangle=1.095$ at
$s/d=4$.  For the forced convection and  the aiding buoyancy cases, $Ri\ge 0$,
the time averaged drag coefficient increases linearly: 
\begin{equation} \langle C_d \rangle = 18.126 \, Ri + 3.288 \qquad Ri \ge 0.
\end{equation} 
On the other hand, in fig.~\ref{fig:sigmaCd-Ri} the standard deviation of the
drag coefficient shows an increase of its oscillating amplitude in opposing
buoyancy cases. The aiding buoyancy suppresses the drag coefficient
oscillation, as described in the qualitative description of the temperature
distribution in the previous section. A sudden transition, at $Ri=-1$, changing
the spacing between the cylinders is found. The decreasing of the spacing from
$4.0$ to $3.6$ causes a limiting effect on the oscillation amplitude of $C_d$.
The standard deviation of the drag force coefficient ($\sigma(C_d)$) of the
array has been detailed on each cylinder in fig.~\ref{fig:sigmaCd-Ri-sxx}. It
is worth to note that all the cylinders are involved in the transition at $Ri =
-1$.
This result is strictly connected with the flow pattern transition depicted in
fig.~\ref{fig:T-contour-neg}.  The standard deviation of the lift coefficient,
$\sigma(C_l)$, confirms the transition at $Ri=-1$ (see
fig.~\ref{fig:sigmaCl-Ri}). Moreover, the temperature distribution in the
near field reveals that for $s/d=3.6$ the boundary layer separations at the
cylinder surfaces are reduced (see fig.~\ref{fig:T-contour-neg-zoom}-a), compared
to the $s/d=4$ spacing (see fig.~\ref{fig:T-contour-neg-zoom}-e), inducing a
dumping in the oscillation amplitude of $C_d$ and $C_l$. At $Ri=0$ and
$Ri=0.25$ a difference in the value of $\sigma(C_l)$ can be recognized. Indeed
it is related to the wake oscillation of the $s/d=4.0$ case with respect to the
stable configuration of $s/d=3.6$ remarking the qualitative discussion about
the instantaneous temperature distribution reported in
section~\ref{subsec:T-distribution}.  $\sigma(C_l)$ is more sensitive than
$\sigma(C_d)$ to the wake oscillation in the near field. In figure
~\ref{fig:St-Ri} the Strouhal number ($St=f d/U^\star$, where $f$ is the
oscillation frequency) of the maximum peak of the fft of the $C_l$ for the
first cylinder is reported. The higher differences in the oscillation frequency
correspond to the cases at $Ri=-1$ and $Ri=0$, where different temperature
patterns have been recognized changing the spacing ratio between the cylinders.
At $Ri=-1$ the sudden jump of $St$ is related to the above mentioned boundary
layer separation reduction that influences the shedding behaviour at $s/d=3.6$
compared to the case at $s/d=4$.  

\subsection{Heat exchange performance}

\begin{figure} \includegraphics[width=0.99\linewidth]{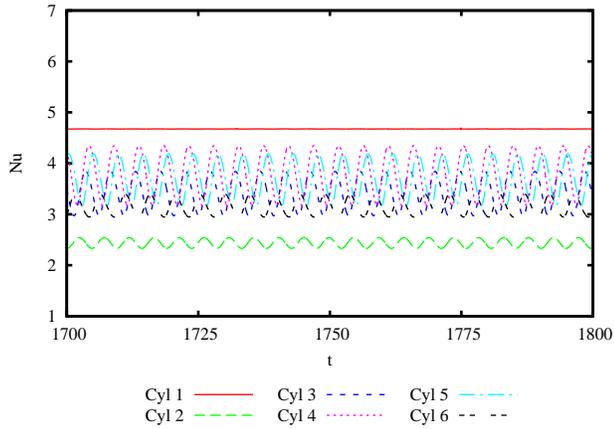}
\caption{Surface averaged Nusselt number ($Nu$) of each cylinder with respect
to the dimensionless time $t$, for $s/d=3.6$ and
$Ri=0$.}\label{fig:Nu-ts-Ri0-s36} \end{figure}

\begin{figure} \includegraphics[width=0.99\linewidth]{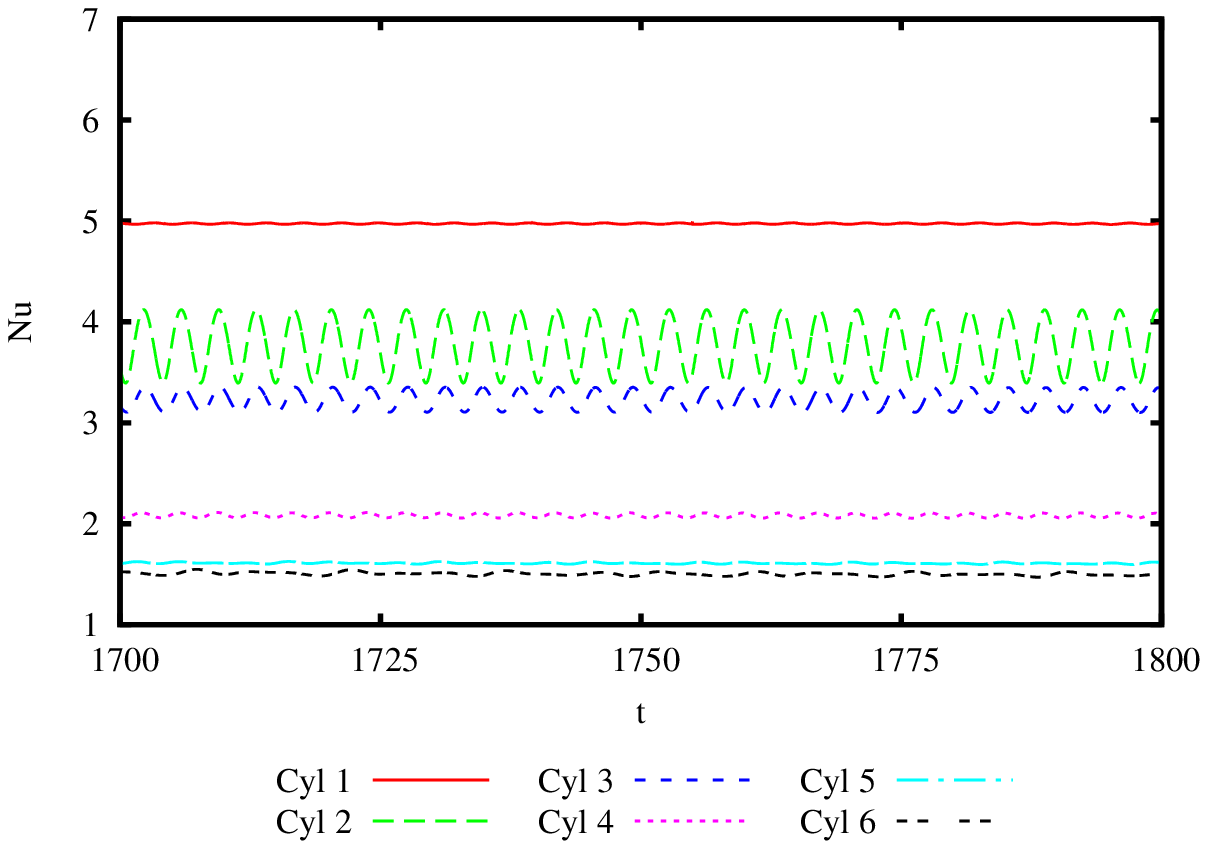}
\caption{Surface averaged Nusselt number ($Nu$) of each cylinder with respect
to the dimensionless time $t$, for $s/d=4.0$ and
$Ri=0$.}\label{fig:Nu-ts-Ri0-s40} \end{figure}

\begin{figure}
\includegraphics[width=0.99\linewidth]{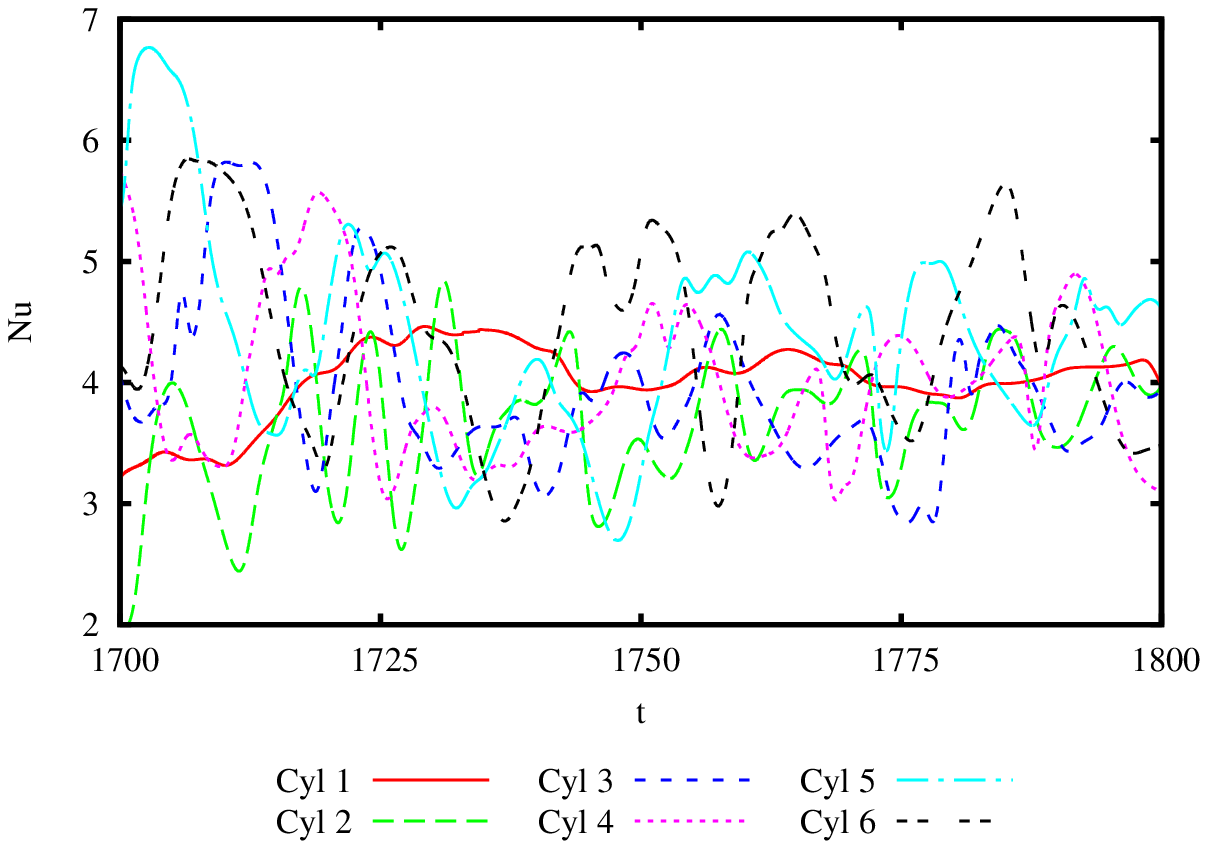}
\caption{Surface averaged Nusselt number ($Nu$) of each cylinder with respect
to the dimensionless time $t$, for $s/d=3.6$ and
$Ri=-1$.}\label{fig:Nu-ts-Ri-1-s36} \end{figure}

\begin{figure}
\includegraphics[width=0.99\linewidth]{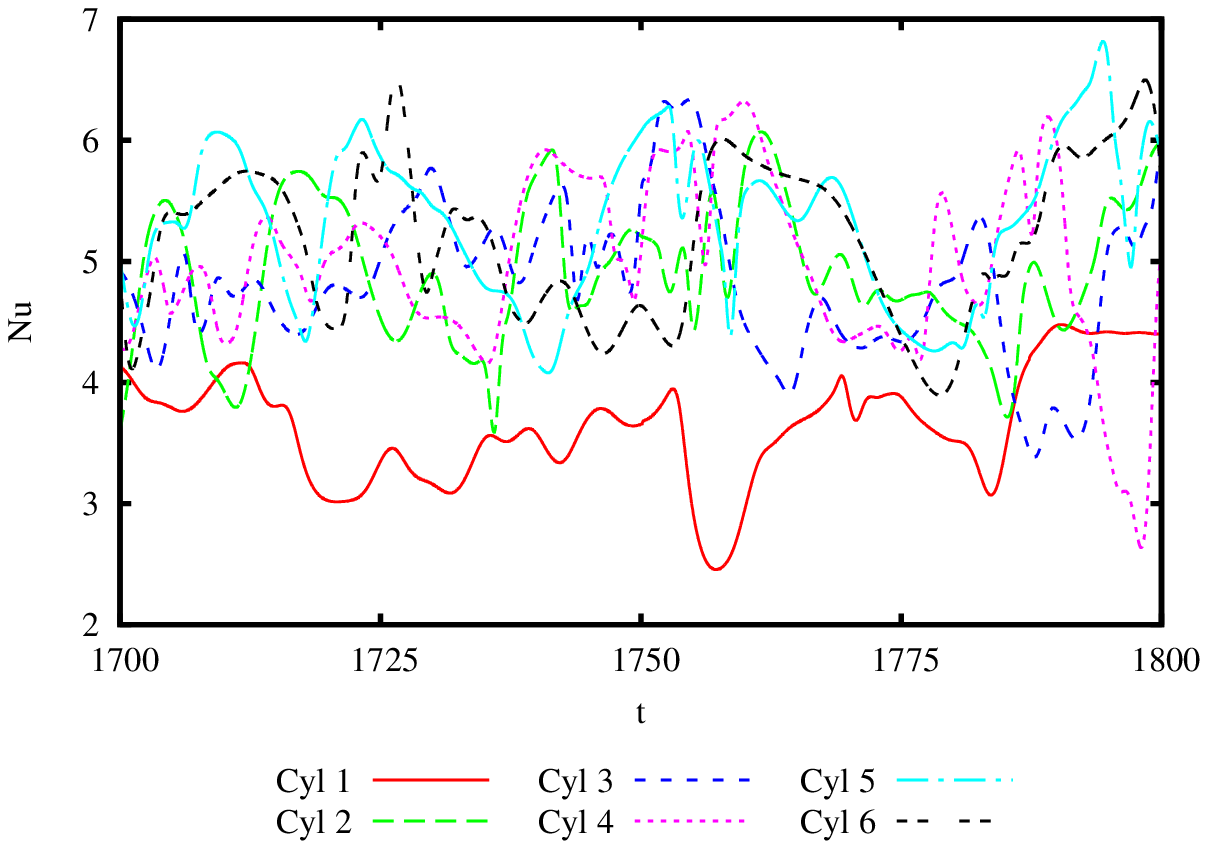}
\caption{Surface averaged Nusselt number ($Nu$) of each cylinder with respect
to the dimensionless time $t$, for $s/d=4.0$ and
$Ri=-1$.}\label{fig:Nu-ts-Ri-1-s40} \end{figure}

\begin{figure} \includegraphics[width=0.99\linewidth]{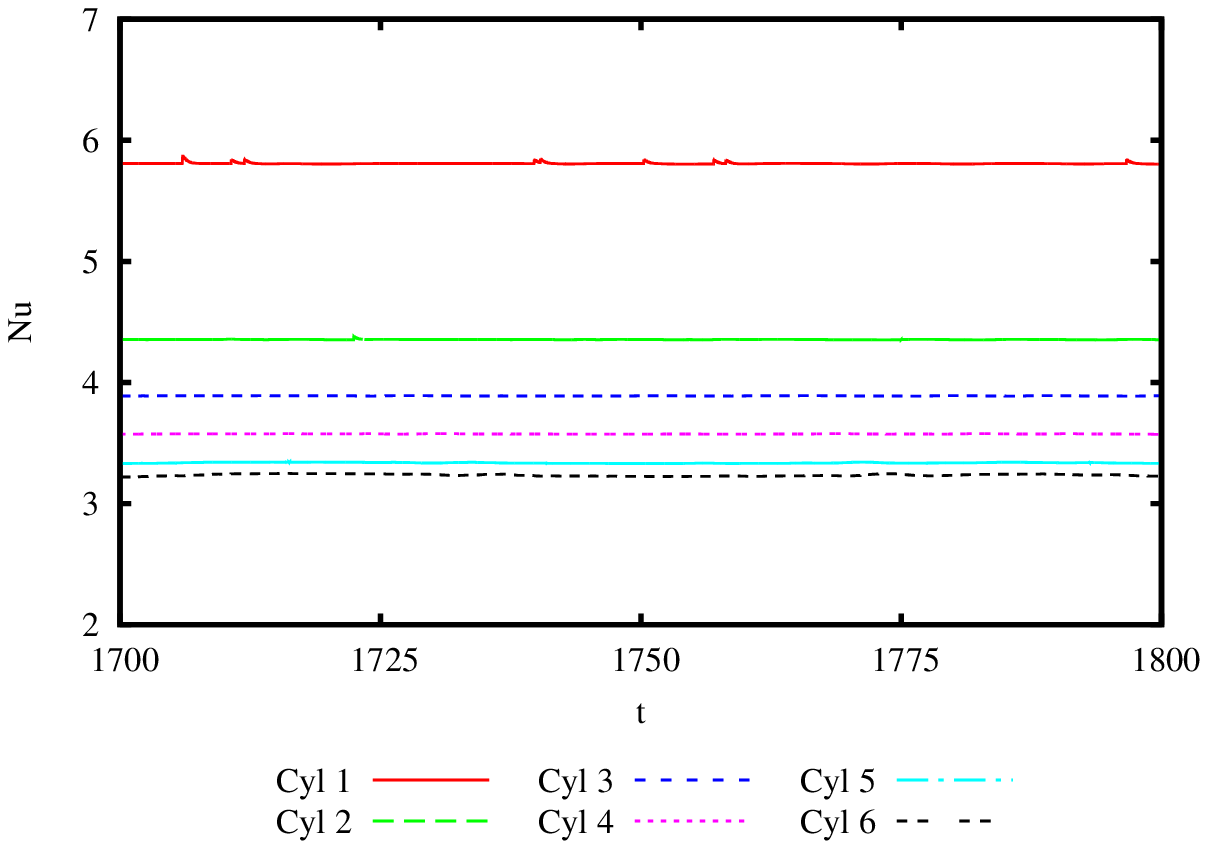}
\caption{Surface averaged Nusselt number ($Nu$) of each cylinder with respect
to the dimensionless time $t$, for $s/d=3.6$ and
$Ri=1$.}\label{fig:Nu-ts-Ri1-s36} \end{figure}

\begin{figure} \includegraphics[width=0.99\linewidth]{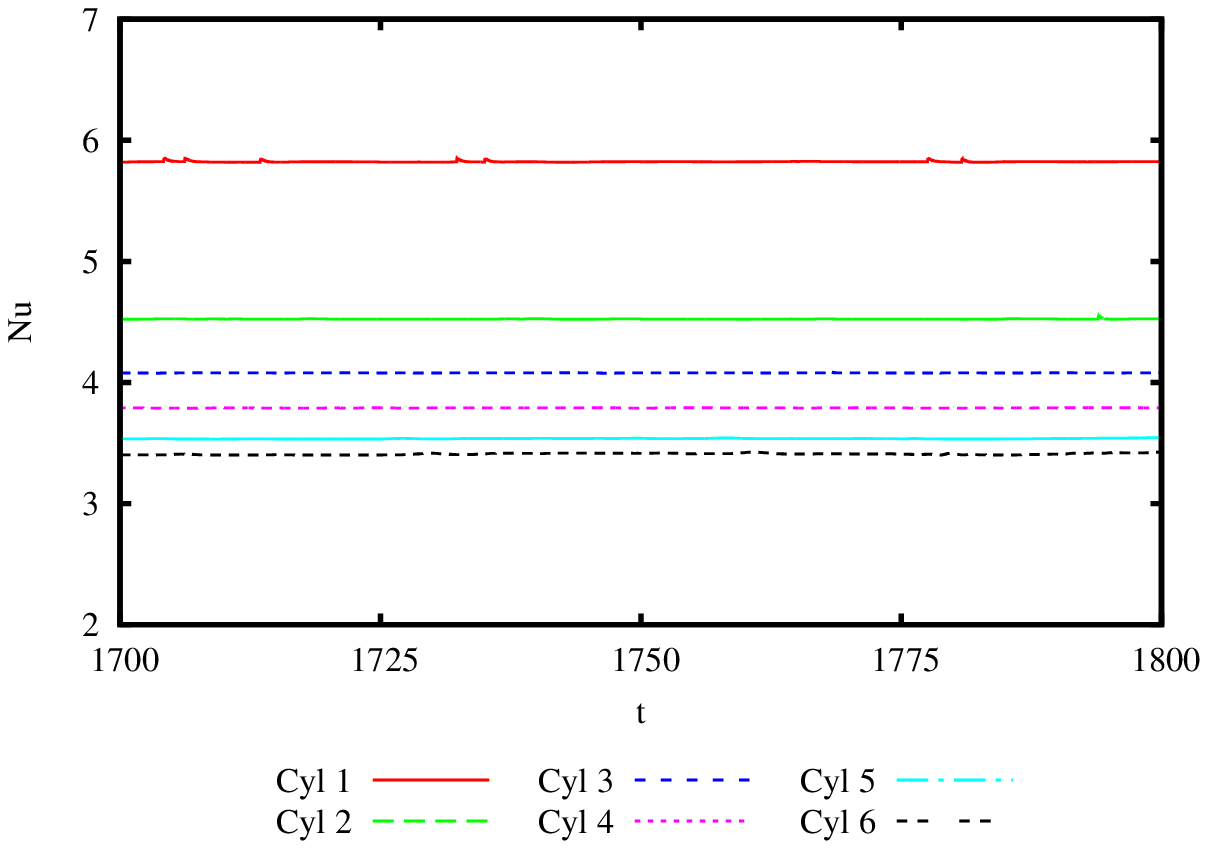}
\caption{Surface averaged Nusselt number ($Nu$) of each cylinder with respect
to the dimensionless time $t$, for $s/d=4.0$ and
$Ri=1$.}\label{fig:Nu-ts-Ri1-s40} \end{figure}

\begin{figure} \includegraphics[width=0.99\linewidth]{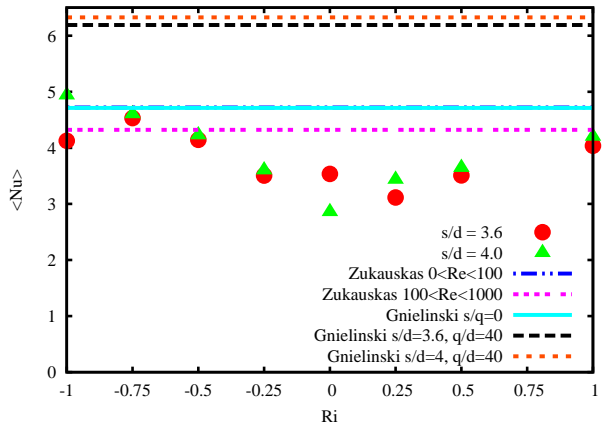}
\caption{Time-averaged Nusselt number ($\langle Nu \rangle$) of the array
with respect to the Richardson number ($Ri$) for $s/d=3.6$ and $s/d=4.0$. The
predictions of Zukauskas~\cite{Zukauskas197293} in two different Reynolds
ranges, $0 - 100$ and $100 - 1000$; and three values of
Gnielinski~\cite{Gnielinski1975145} for a single line of six cylinders
($s/q=0$), and two configurations of an infinite array of six cylinders
($s/d=3.6 - 4$ and $q/d=40$) are reported.}\label{fig:Nu-Ri} \end{figure}

\begin{figure*}
\includegraphics[width=0.99\linewidth]{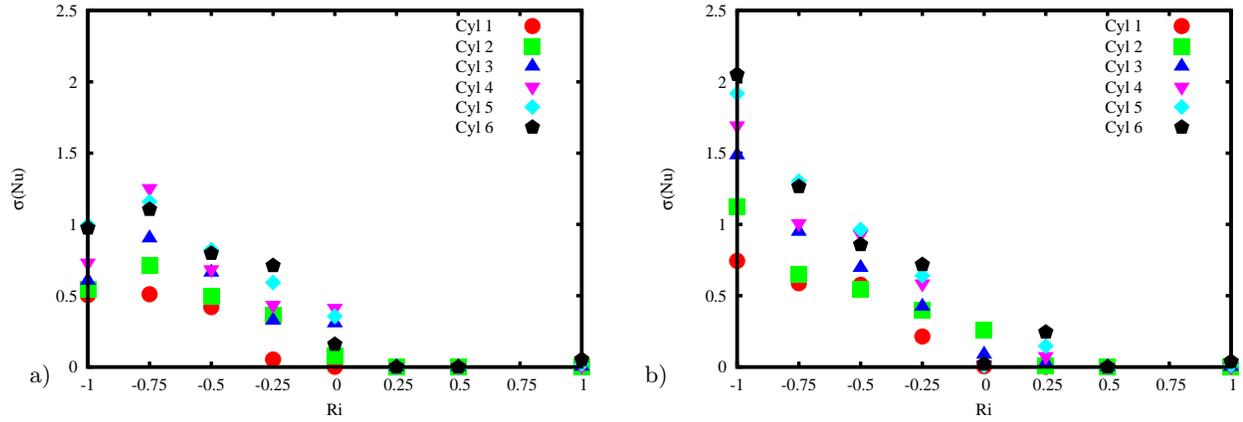}
\caption{Standard deviation of the Nusselt number ($\sigma(Nu)$) of each
cylinder respect to the Richardson number ($Ri$) for $s/d=3.6$ (a) and
$s/d=4.0$ (b).}\label{fig:sigmaNu-Ri-sxx} \end{figure*}

The thermal performance of the cylinders array with respect to the buoyancy
force modulation is evaluated by means of the Nusselt number. The
instantaneous surface averaged Nusselt numbers, $Nu$, for each cylinder have
been sampled during the simulations. It is obtained averaging the local Nusselt
number, $Nu'=\partial T/\partial n$. The forced convection case ($Ri=0$) shows
a monochromatic response of the surface averaged Nusselt number for each
cylinder for both the spacing ratio, $s/d=3.6$ (fig.\ref{fig:Nu-ts-Ri0-s36})
and $s/d=4.0$ (fig.\ref{fig:Nu-ts-Ri0-s40}).  The opposing buoyancy ($Ri=-1$)
induces more complex oscillations of $Nu$ as reported in fig.~\ref{fig:Nu-ts-Ri-1-s36} for $s/d=3.6$ and
in fig.\ref{fig:Nu-ts-Ri-1-s40} for $s/d=4.0$. On the other hand, the aiding
buoyancy, for $Ri=1$, suppresses the oscillations on all the cylinders as
reported in fig.~\ref{fig:Nu-ts-Ri1-s36} for $s/d=3.6$ and in
fig.~\ref{fig:Nu-ts-Ri1-s40} for $s/d=4.0$.  Thus, the time averaged Nusselt
number, $\langle Nu \rangle$, and its standard deviation, $\sigma(Nu)$, have
been extracted.  First, the time-average Nusselt number of the entire array is
analyzed (see fig.~\ref{fig:Nu-Ri}).  
The results are compared to the prediction of time-averaged heat exchange in
case of tube banks of Zukauskas~\cite{Zukauskas197293}:
\begin{equation}\label{eq:Zukauskas} Nu= C_1 C_2 Re^m Pr^n \end{equation}
where $C_1$, $m$ and $n$ are related to the Reynolds number, whereas $C_2$ is a
correction factor for an array of less then $20$ cylinders. The values of the
coefficients here considered are reported in table~\ref{tab:zukauskas}. The
influence of longitudinal and transversal spacing between the cylinders are not
included in the model.  
\begin{table} \centering \caption{Parameters used in  eq.~\ref{eq:Zukauskas}
according to the Zukauskas~\cite{Zukauskas197293} model for the prediction of
the Nusselt number of a tube bundle.} \label{tab:zukauskas} \begin{tabular}{c c
c c c} \hline\\ $Re$ & $C_1$ & $C_2$ & $m$ & $n$ \\ \hline\\ $0 - 100$ & $0.9$
& $0.945$ & $0.4$ & $0.36$ \\ $100 - 1000$ & $0.52$ & $0.945$ & $0.5$ &
$0.36$\\ \hline \end{tabular} \end{table}
Gnielinski~\cite{Gnielinski1975145} extends the prediction model of the Nusselt
number for a single cylinder to the tube bundle case  introducing a correction
term ($f_a$) including the influence of longitudinal ($s/d$) and transversal
($q/d$) spacing ratio:
\begin{equation}\label{eq:Gnielinski} Nu=  f_a
\left(0.3+\sqrt{Nu_{lam}^2+Nu_{turb}^2}\right) \end{equation}
where
\begin{align} Nu_{lam}&=&0.664 Re_\lambda^{0.5} Pr^{0.33},\nonumber\\
Nu_{turb}&=&\frac{0.037 Re_\lambda^{0.8}Pr}{1+2.443
Re_\lambda^{-0.1}(Pr^{2/3}-1)}, \nonumber \end{align} \begin{equation}
Re_\lambda=\frac{\pi}{2} Re, \qquad
f_a=1+\left[\frac{0.7(\frac{s/d}{q/d}-0.3)}{\psi^{1.5}\left(\frac{s/d}{q/d}+0.7\right)^2}\right].\nonumber
\end{equation}
Imposing $s/q\rightarrow0$ and the void ratio $\psi=1-\pi/(4 q/d)\simeq 1$, a
single line of cylinders can be approximated. In these hypotheses, the
Gnielinski~\cite{Gnielinski1975145} model remarks the result of
Zukauskas~\cite{Zukauskas197293} in the range $0<Re<100$.  The model is very
sensitive to the transversal spacing, i.e., considering $q/d=40$ in the
Gnielinski~\cite{Gnielinski1975145} model, according to the influence of
multiple lines, the Nusselt number increases up to $6.190$ and $6.326$ for
$s/d=3.6$ and $s/d=4$, respectively.  According to the numerical simulations,
$Ri$ and $s/d$ influence the averaged Nusselt number that the simplified models
overestimate. The higher spacing between the cylinders gives rise to a higher
Nusselt number in case of aiding or opposing buoyancy. However, in case of
forced convection the tighter spacing, $s/d = 3.6$, leads to a higher heat
exchange.  According to Fornarelli et al.~\cite{FornarelliOresta}, such
behaviour is related to an enhanced cold fluid entrainment in the gap between
two successive cylinders for the $s/d=3.6$ case, due to the near field flow
oscillation (see fig.~\ref{fig:timeseries-Ri0}).  It is worth to note that the
minimum $Nu$ for $s/d=3.6$ and $s/d=4$ corresponds to $Ri=0.25$ and $Ri=0$
respectively. The increase of opposing or aiding buoyancy force induces a
quasi-symmetrical heat transfer enhancement.  $Ri=-0.75$ case is well
approximated by the Zukauskas~\cite{Zukauskas197293} and
Gnielinski~\cite{Gnielinski1975145} models for both spacings.  $Ri=-1$ induces
a different behaviour for the two spacings here considered.  Thus, with respect
to $Ri=-0.75$, the tighter configuration ($s/d=3.6$) shows a reduction of
Nusselt number. For $s/d=4$, at $Ri=-1$ the Nusselt number reaches its maximum
($Nu=4.94$).
\begin{figure} \includegraphics[width=0.99\linewidth]{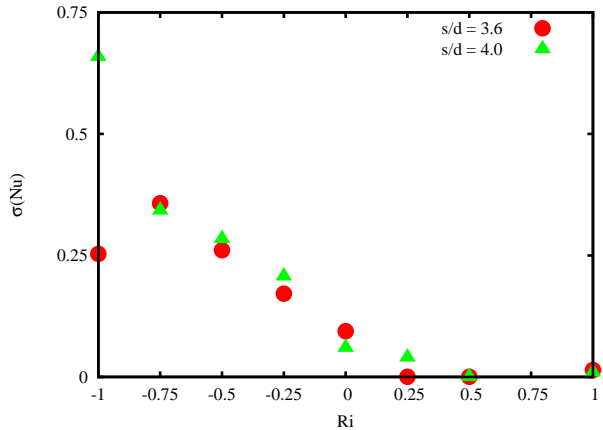}
\caption{Standard deviation of the Nusselt number ($\sigma(Nu)$) of the array
with respect to the Richardson number ($Ri$) for $s/d=3.6$ and
$s/d=4.0$.}\label{fig:sigmaNu-Ri} \end{figure}
Figure~\ref{fig:sigmaNu-Ri} shows the amplitude of the $Nu$ oscillation with
respect to the $Ri$ and the spacing ratio. The buoyancy effect on the flow
pattern influences the amplitude of the averaged Nusselt number. The behaviour
of the standard deviation of the Nusselt number shows the stabilization effect
of the aiding buoyancy except for $s/d=4$ and $Ri=0.25$.  The wake oscillation
depicted in the instantaneous temperature distribution for $s/d=4$ and
$Ri=0.25$, reported in fig.~\ref{fig:T-contour-pos}-f, causes the onset of
Nusselt amplitude even in the case of aiding buoyancy. The opposing buoyancy
increases the oscillation amplitude of the heat exchange. From $Ri=-0.25$ to
$Ri=-0.75$ a linear increase of the standard deviation of $Nu$ for both
spacings has been found. The small differences of $\sigma(Nu)$ for the two spacings are
influenced by the wider gap between the cylinders of the $s/d=4$ case. The
further increase of opposing buoyancy force, $Ri=-1$ shows a transition of the
oscillation amplitude of Nusselt number for the two spacing. The $s/d=4$,
$Ri=-1$ case holds an increase of the amplitude of $Nu$ reaching
$\sigma(Nu)=0.66$, otherwise at $s/d=3.6$ $\sigma(Nu)=0.25$. The $\sigma(Nu)$
of each cylinder is reported in fig.~\ref{fig:sigmaNu-Ri-sxx}. At $Ri=-1$ the
spacing ratio increase produces higher oscillations of Nusselt number on each
cylinder. At $Ri=-0.25$ the spacing increase enhances the $\sigma(Nu)$ of the
first cylinder. At $Ri=0.25$ the higher value of the averaged $\sigma(Nu)$
reported in fig.~\ref{fig:sigmaNu-Ri} at $s/d=4$ is mainly related to the
oscillating amplitude of the trailing cylinders.

\section{Conclusions}

We can infer, by means of numerical simulations, that the buoyancy force and the
spacing between the cylinders, arranged in a single six elements row, affect their
performance in terms of forces and heat exchange. Generally, the aiding or the
opposing buoyancy induces a stable or unstable behaviour of the flow,
respectively. However, in case of opposing buoyancy a flow transition has been
recognized. In case of aiding buoyancy the spacing influences the secondary
instability of the wake in the far field, therefore the differences in terms of
force and heat exchange between the fluid and the cylinders are small because
they are linked with the near field behaviour. On the other hand, in case of
opposing buoyancy the flow instabilities in the near field affect the array
performance. The oscillation amplitudes of $C_d$, $C_l$ and $Nu$ become
relevant with respect to the mean quantities for both the spacing here investigated.
At $Ri = -1$ the spacing affects heavily the oscillation amplitude of the
measured quantities.  At $Ri = -1$ and $s/d = 4$ the standard deviation of the
performance coefficients, $\sigma(C_d)$, $\sigma(C_l)$ and $\sigma(Nu)$
increases with $Ri$. However, for a spacing ratio $s/d=3.6$, even with
a strong opposing buoyancy ($Ri=-1$), the flow is able to rearrange itself in a
more ordered wake pattern configuration limiting the oscillation amplitude of
the performance coefficients. In the range of parameters here presented, the
heat exchange and force coefficients of the cylinders array appear very
sensitive with respect to the value of Richardson number. Thus, this work highlights
the care that should be taken in using existing predictive models to estimate the
heat transfer around a finite number of circular cylinders in case of mixed
convection. 

\begin{acknowledgment}
The authors would like to thank the IT staff of ``Centro Cultura Innovativa
d'Impresa'' of ``University of Salento'', where the simulations were carried out,
for their technical help.
\end{acknowledgment}

%
% ...

\bibliographystyle{asmems4}

\bibliography{biblio}


\end{document}